# Cloud Security Challenges: Investigating Policies, Standards, and Guidelines in a Fortune 500 Organization

George Grispos, School of Computing Science, University of Glasgow, Glasgow, United Kingdom, g.grispos.1@research.gla.ac.uk

William Bradley Glisson, Humanities Advanced Technology and Information Institute, University of Glasgow, Glasgow, United Kingdom, Brad.Glisson@glasgow.ac.uk

Tim Storer, School of Computing Science, University of Glasgow, Glasgow, United Kingdom, timothy.storer@glasgow.ac.uk

## Abstract[1]

Cloud computing is quickly becoming pervasive in today's globally integrated networks. The cloud offers organizations opportunities to potentially deploy software and data solutions that are accessible through numerous mechanisms, in a multitude of settings, at a reduced cost with increased reliability and scalability. The increasingly pervasive and ubiquitous nature of the cloud creates an environment that is potentially conducive to security risks. While previous discussions have focused on security and privacy issues in the cloud from the end-users perspective, minimal empirical research has been conducted from the perspective of a corporate environment case study. This paper presents the results of an initial case study identifying real-world information security documentation issues for a Global Fortune 500 organization, should the organization decide to implement cloud computing services in the future. The paper demonstrates the importance of auditing policies, standards and guidelines applicable to cloud computing environments along with highlighting potential corporate concerns. The results from this case study has revealed that from the 1123 'relevant' statements found in the organization's security documentation, 175 statements were considered to be 'inadequate' for cloud computing. Furthermore, the paper provides a foundation for future analysis and research regarding implementation concerns for corporate cloud computing applications and services.

**Keywords: Cloud Computing, Information Security, Corporate Policy, Requirements, Standard, Guidelines**

## 1. Introduction

Cloud computing is rapidly becoming pervasive in today's globally integrated networks. This emerging paradigm allows an organization to reduce costs and develop more reliable and highly scalable solutions [1]. There is an apparent consensus that cloud computing is increasingly integrating into the business environment [2, 3]. The 2011 Ernst and Young Global Information Security Survey [3] reported that 36% of respondents were currently using cloud computing services, as compared to 23% from the previous year. Furthermore, this report goes on to state that 25% of respondents are currently evaluating or investigating the use of cloud services in the subsequent year. In concert with these findings, the AMD Global Cloud Computing Study [2] reported that 37% of surveyed businesses globally are deploying applications or storing data in cloud environments. Cloud computing services are also being utilized by governments in an attempt to reduce costs and improve the efficiency of IT solutions within their agencies [4]. The United States Federal Government is expected to save $10 billion in IT costs by migrating government services to cloud computing [5].

---

[1]**Please cite this paper as:** Grispos, G., Glisson, W.B., and Storer, T. (2013) *Cloud Security Challenges: Investigating Policies, Standards, and Guidelines in a Fortune 500 Organization.* In: 21st European Conference on Information Systems, 5-8 Jun 2013, Utrecht, The Netherlands.

Although the benefits of cloud computing are attractive, a major concern is the security of data and applications in these environments. This apprehension focuses on data confidentiality and auditability, data lock-in, a lack of encryption, the impact of borderless and virtualized environments, as well as cloud providers not addressing requirements from a business, information technology or security perspective [1, 3]. These fears are being validated as cloud adopters witness an increase in the number of cloud infrastructure security incidents as compared to traditional IT infrastructure security events [6]. The real-world impact is that security incidents and data breaches can have financial consequences on a corporate organization [7]. The Ponemon Institute [8] indicates that although the cost of data breaches has declined in the past year, data breaches are still reported to have cost British and German organizations on average between $2.7 million and $4.4 million. In addition to economic losses, security incidents and data breaches can result in a loss of customers, damaged reputations, delayed software releases and a reduction in investor confidence [7]. To reduce the likelihood and impact of security incidents and data breaches, organizations implement information security programs [9]. The primary objective of an information security program is to protect the integrity, confidentiality and availability of information and applications while ensuring that legal and regulatory requirements are also fulfilled [9, 10]. Hence, an organization's overall strategy towards specific environments like private and public clouds should be reflected in their policies, standards and guidelines. Although the European Network and Information Security Agency [11] and the Cloud Security Alliance [12] have defined security documentation for cloud computing environments, very little research investigates the application of this documentation towards mitigating cloud-related security incidents and data breaches.

This research presents an empirical case study investigating information security policies, standards, and guidelines in a Global Fortune 500 organization. The research begins to identify potential areas of information security documentation exposure from the perspective of a large organization through the examination of 164 policies, 1,255 standards and 674 guidelines. The research contribution is an empirical report identifying security issues along with research opportunities in policies, guidelines and procedures in large organizations that are considering migrating to a public cloud environment. The structure of the paper is as follows. Section 2 presents an overview of cloud computing security challenges and issues identified from the literature, Section 3 defines the method of data collection implemented in this study. Section 4 provides insight into the organization through an analysis of existing documentation and identifies current enhancement opportunities. Section 5 draws conclusions from the work conducted and presents ideas for further work.

## 2. Analysis of Cloud Security Issues

Cloud computing has been defined as "a model for enabling ubiquitous, convenient, on-demand network access to a shared pool of configurable computing resources (e.g., networks, servers, storage, applications, and services) that can be rapidly provisioned and released with minimal management effort or service provider interaction" [13]. Grispos et al., [14] have discussed the different service and deployment configuration options of a cloud computing environment from an information security perspective.

The security of data and applications is a concern for organizations wishing to adopt cloud computing services. Armbrust et al., [1] cite a number of security concerns regarding cloud adoption including data confidentiality and auditability, business continuity and data lock-in. Subashini & Kavitha [15] highlighted security issues applicable to various layers of the cloud computing environment while noting that security needs will vary for each delivery model. The Cloud Security Alliance (CSA) has also identified several critical areas in cloud environments [12]. The high-level domains along with associated research in those areas are available in Table 1 – Cloud Security Research. Initial areas of interest are unauthorized data access, vendor lock-in, legal issues and virtualization.

One of the biggest concerns for an organization, considering the adoption of cloud computing, is preventing unauthorized access to data resources [16]. This is a concern because an organization's traditional authentication and authorization framework does not naturally extend into the cloud environment and requires modification to support cloud computing services [17]. This problem is particularly evident in Software-as-a-Service (SaaS) environments, where a user's credentials are likely to be stored in the SaaS providers' databases, and who effectively become responsible for security and access control mechanisms [18]. Having different authentication systems for internal and cloud-based resources is likely to further complicate the issue and become impractical over time [16]. One proposed solution is the use of a federated identity management system, allowing users to access the cloud environment from anywhere outside the traditional workplace while authentication credentials remain within the corporate boundary [19]. Additionally, Kaufman [20] suggests that to ensure data confidentiality, integrity, and availability, cloud providers must offer capabilities that include the encryption of data across their entire infrastructure and a backup mechanism to replace corrupted data. A regular backup mechanism can ensure a quick recovery of service in the case of disaster or accidental data loss [17, 21]. The Data Protection Working Party [22] recommends that organizations utilizing cloud services instruct cloud providers to only store applications, primary and backup data in pre-defined physical locations. The storage of data in known physical locations also supports compliance with regulatory and jurisdictional requirements [12, 22].

Another concern is that migration from one cloud provider to another can lead to issues like data remembrance and vendor lock-in [12, 23]. Currently, immediate compliance with a customer's request to comprehensively destroy data from the provider's infrastructure is uncertain [23]. For example, Google's policy is that once data is deleted, it is removed from all active and replication servers. However, the data is only completely removed when it is overwritten by another customer's data over a period of time [24]. This brings up an interesting question. When a customer decides that they would like to leave a service, what confirmation can cloud providers offer to verify that all of the user's data has been completely erased from their storage facility? Moyle [25] has hypothesized that cloud service providers will restrict their customers logical access to their physical data through a number of 'security controls'. As a result, this can lead to the phenomenon referred to as 'vendor lock-in' [23]. Vendor lock-in can inhibit an organization from migrating from one provider to another [1, 12]. This environment can also be encouraged thorough the implementation of proprietary or non-standard Application Programming Interfaces (APIs) and data storage methods.

Large organizations are often familiar with legislative compliance requirements for logging system activity. Often legal and regulatory requirements such as the Sarbanes-Oxley (SOX) Act of 2002 and the Health Insurance Portability and Accountability Act (HIPAA) of 1996 require that an accurate time source is used to synchronize the system clocks of any potential logs [26]. The challenge arises in the cloud environment where data and log files are stored in more than one geographical location and possibly located in different time zones. Cloud providers need to ensure that a universal time source is used across its infrastructure to eliminate potential discrepancies in timestamps [12]. Further challenges identified include provisions on how long logs must be maintained by the cloud provider, where the cloud provider can store logs, who will be responsible for the creation and maintenance of the logs, what security controls must be in place to prevent unauthorized modification, as well as under what conditions the logs can be deleted [27-29].

Virtualization is also expected to present a number of security challenges in cloud environments [16, 30-33]. Ibrahim, et al., [31] discussed how known security issues, within virtualized environments, can also be found in the various components of an Infrastructure-as-a-Service (IaaS) implementation. This issue is further enhanced in a multi-tenancy virtual machine-based infrastructure where, in the event malicious code is inserted into the virtualization layer, an attacker can gain full control over the physical infrastructure and all the virtual machines connected to that particular infrastructure [16, 30]. Ristenpart et al., [32] have successfully demonstrated that this practice is a security concern for both the cloud provider and the organizations using the resulting service. They demonstrated how cross-virtual machine attacks can result in data leakage between virtual machines which share the same physical resources in the Amazon EC2 Cloud [32]. Complicating matters, Wei, et al., [33] described how an attacker can modify virtual machine images hosted in the cloud. These virtual machine images are used to create virtual machine instances which could be used to allow an attacker to access the virtual machine, data, applications and execute malicious code [33].

| CSA Domain Name | Cite(s) |
|---|---|
| **Governance and Enterprise Risk Management** | [20, 28, 34, 35] |
| **Legal Issues: Contracts and Electronic Discovery** | [14, 15, 36-38] |
| **Compliance and Audit Management** | [1, 15, 20, 34, 39] |
| **Information Management and Data Security** | [15-17, 20, 40, 41] |
| **Interoperability and Portability** | [17, 23, 40] |
| **Traditional Security, Business Continuity and Disaster Recovery** | [15, 34, 40] |
| **Data Centre Operations** | [12, 28] |
| **Incident Response** | [14, 36, 42, 43] |
| **Application Security** | [17, 20, 34, 40, 41] |
| **Encryption and Key Management** | [12, 39] |
| **Identity, Entitlement and Access Management** | [15-17, 41] |
| **Virtualization** | [16, 24, 32, 33, 44] |

*Table 1 - Cloud Security Research*

Cloud security is a multi-faceted environment that consists of the provider, the user, and any other third-party vendor who provides security software and services to either the provider or the user [40]. Research has focused on a number of security issues in cloud computing environments. However, very little empirical research has been conducted to examine real-world information security documentation in an organization assessing cloud computing 'security-readiness'.

## 3. Method

Based on growing industry interest in cloud computing environments, it was hypothesized that current information security documents are potentially putting organizations at risk. Hence, this research empirically investigates security-readiness through exploratory interviews and an examination of information security documents relevant to cloud computing environments. Exploratory case studies provide insight into understanding real-world problems and context along with identifying future areas of research [45]. The case study was completed between May and August 2012 and focuses on the information security documentation implemented in a Global Fortune 500 organization. In order to attain accurate information and foster an environment where all parties are comfortable presenting commercially sensitive information, the name of the organization is being withheld to ensure organizational anonymity. Hence, the names of the documents, security processes and groups have all been altered and the results of the exploratory interviews conducted with employees are presented anonymously.

An analysis of the existing literature was conducted to identify and develop a broad understanding of the major themes regarding information security concerns in the cloud. The literature was collected from a number of sources including academic papers from online collections, industrial whitepapers and relevant textbooks. For the purpose of this discussion, the authors highlight five domains from the Cloud Security Alliance report [12] and relevant literature. These domains include: Information Management and Data Security, Interoperability and Portability, Incident Response, Encryption and Key Management, and Virtualization.

In order to acquire an understanding of the Global Fortune 500 organization's context, exploratory interviews were conducted with four senior members of the organization's security team. These senior members included: two security analysts, an enterprise security architect and an IT security manager. The security team is responsible for enforcing security controls during the application development process, as well as implementing everyday operational security. The purpose of the exploratory interviews was to identify the high-level information security documentation containers used within the organization and to investigate what security documents are applied to various projects. Each interview lasted 45 minutes and the interviewee's responses were recorded by hand. All hand-written notes were digitally recorded approximately within one hour after the interview was completed. The security documentation containers identified in the interviews are as follows:

- The *Technology Risk Repository (TRR)* contains documentation describing appropriate communication within the organization. These documents specify that all confidential or sensitive information must be encrypted, the identity of recipients of confidential or sensitive information must be verified and that all devices must be monitored for inappropriate use.
- The *Information Security Repository (ISR)* contains documentation providing high-level requirements regarding the organization's policy framework and compliance infrastructure. This includes information pertaining to communications and operations management, specific cryptographic algorithms, incident management procedures and business continuity requirements.
- The *Risk Management Repository (RMR)* contains documentation which defines how the organization communicates with external contractors, customers and regulators. It also contains policies to ensure compliance with legal and regulatory requirements.

For the purpose of this research, three main types of security documents were identified from the repositories: policies, guidelines and standards. The objective of an information security policy is to define "what" is to be done, while standards and guidelines specify "how" the policy objectives will be implemented [9, 46]. Information security policy statements define "the roles and responsibilities of the employees to safeguard the information and technology resources of their organizations" and are high-level statements which are technology and solution-independent documentation [9, 47]. Standards and guidelines provide detailed information enabling employees to achieve objectives stated in security policies [48]. The literature review was used as a filtering guide to identify areas that present cloud security issues.

## 4. Security Documentation Analysis

A total of 127 security documents containing 2,093 security-related statements were available for examination. The Technology Risk Repository (TRR) consisted of 89 security documents containing 1,268 statements. The Information Security Repository (ISR) contained 14 security documents containing 232 statements. The Risk Management Repository (RMR) consisted of 24 security documents containing 593 statements. Statements within the various security documents were categorized as either 'relevant', 'non-relevant' or 'further research required'. A 'relevant' classification is one which, in its current format, is applicable to cloud computing environments. For example, a statement like 'cryptography mechanisms must be utilized to achieve data integrity and confidentiality' would also apply to any potential cloud computing service used by the organization. A 'non-relevant' classification is one which is not applicable to cloud computing environments. An example of such a statement is a 'clear desk' policy which dictates that all employees must not leave sensitive information on their desks overnight. A classification of 'further research required' indicates that there is the potential for the statement to be relevant but additional investigations are required. Such an example would be specific server-side configurations for UNIX and Windows networked machines. This is due to the fact that it is not clear if the specific configurations will be applicable to, or can be implemented in, cloud-based instances. This information is available in Table 2 - Repository Analysis.

| Document Name/Repository | TRR | ISR | RMR | Total |
|---|---|---|---|---|
| *Total Policy Documents* | 56 | 5 | 0 | 61 |
| *Total Policy Statements* | 56 | 108 | 0 | 164 |
| *Relevant Policy Statements* | 34 | 81 | 0 | 115 |
| *Non- Relevant Policy Statements* | 22 | 27 | 0 | 49 |
| *Require Further Research Policy Statements* | 0 | 0 | 0 | 0 |
| *Total Standard Documents* | 27 | 4 | 0 | 31 |
| *Total Standard Statements* | 1194 | 61 | 0 | 1255 |
| *Relevant Standard Statements* | 595 | 28 | 0 | 623 |
| *Non- Relevant Standard Statements* | 101 | 33 | 0 | 134 |
| *Require Further Research Standard Statements* | 498 | 0 | 0 | 498 |
| *Total Guideline Documents* | 6 | 5 | 24 | 35 |
| *Total Guideline Statements* | 18 | 63 | 593 | 674 |
| *Relevant Guideline Statements* | 1 | 62 | 322 | 385 |
| *Non- Relevant Guideline Statements* | 17 | 1 | 24 | 42 |
| *Require Further Research Guideline Statements* | 0 | 0 | 247 | 247 |

*Table 2 - Repository Analysis*

Additional analysis categorized 'relevant' statements as being 'adequate' or 'inadequate'. An 'adequate' statement addresses cloud security risks for any potential cloud computing services used by the organization. For example, a statement like 'all legal and regulatory requirements must be considered when storing personal data' would be considered as an 'adequate' statement. An 'inadequate' statement does not mitigate cloud security risks. An example of such a statement is 'all virtual machine instances must be locked down using the VMware Lockdown Standard. This statement was considered 'inadequate' as the standard assumes the analyst 'locking downing' the virtual machine has control over the hypervisor. This assumption could be invalid in a public cloud environment. From the 115 'relevant' policy statements, eight were considered to be 'inadequate' for cloud computing. Out of the 623 'relevant' standard statements, 61 were considered to be 'inadequate' for cloud computing. Finally, from the 'relevant' 385 guideline statements, 106 were considered to be 'inadequate' for cloud computing. This information is summarized in Table 3 - Relevant Analysis. The results indicate that opportunities exist within the organization to improve security documents related to Interoperability and Portability, Information Management and Data Security, Incident Response, Encryption and Key Management, and Virtualization.

## 4.1 Interoperability and Portability

The organization currently does not have a process to govern a migration from one cloud provider to another or a process to govern how to completely remove its data and applications from the infrastructure of a legacy provider after a migration. This lack of a policy process could lead to eventual vendor lock-in. This process would need to define requirements for cloud providers to ensure they are able to guarantee that data migrated from the legacy provider to a new provider is done so in a suitable and readable format. When selecting a cloud provider, the organization must also ensure that the application programming interfaces (APIs) used by the cloud provider are either non-proprietary or the provider can guarantee that data or application utilizing these interfaces can be easily migrated or removed from its infrastructure. Additionally, organizations should ensure that virtual machine instances and virtual machine images are also stored in a format which can allow them to be easily migrated between providers, for example the Open Virtualization Format (OVF) [12]. Any future cloud migration process should also define how logs, metadata and backups of data will be migrated from one provider to another. Will the legacy cloud provider perform the physical migration or will the organization need to handle this issue?

| Document Name/Repository | TRR | ISR | RMR | Total |
|---|---|---|---|---|
| *Total Policy Statements* | 56 | 108 | 0 | 164 |
| *Relevant Policy Statements* | 34 | 81 | 0 | 115 |
| *Adequate Policy Statements* | 28 | 79 | 0 | 107 |
| *Inadequate Policy Statements* | 6 | 2 | 0 | 8 |
| *Total Standard Statements* | 1194 | 61 | 0 | 1255 |
| *Relevant Standard Statements* | 595 | 28 | 0 | 623 |
| *Adequate Standard Statements* | 537 | 25 | 0 | 562 |
| *Inadequate Standard Statements* | 58 | 3 | 0 | 61 |
| *Total Guideline Statements* | 18 | 63 | 593 | 674 |
| *Relevant Guideline Statements* | 1 | 62 | 322 | 385 |
| *Adequate Guideline Statements* | 1 | 53 | 225 | 279 |
| *Inadequate Guideline Statements* | 0 | 9 | 97 | 106 |

*Table 3 - Relevant Statements Analysis*

## 4.2 Information Management and Data Security

The Identity Lifecycle Management process document defines the minimum security requirements for the identity lifecycle management processes used in the organization. The purpose of this document is to minimize the potential impact from intentional and unintentional unauthorized access to any of the organization's information resources. The document does not specifically address provisions on how cloud computing will integrate into the organization's current identity management (IdM) process. It does however state a formal process for establishing federated requests from third-party organizations. This third-party process specifies that the appropriate legal, regulatory and cryptographic controls must be in place and that the federating third-party complies with the organization's security policies. The process would be more comprehensive if it included details on the implementation of IdM for Software-as-a-Service (SaaS) applications in the cloud. If the organization performs a security assessment of a cloud provider, security analysts will need to establish how the provider intends to implement IdM. The third-party security assessment documents do not currently request this information. Hence, these documents need to be extended to include such provisions. This would allow the organization to identify the type of IdM solution the provider currently implements and to establish if this solution can be integrated with the organization's current IdM framework.

Documentation within the organization requires third-parties to have controls in place to ensure continuous availability of data which includes performing regular backups. Specific guidelines assume that a cloud provider will be responsible for providing backups. This requirement should be established on a case-by-case basis, as some cloud providers do not provide a full backup facility as part of their Service Level Agreement (SLA). For example, Amazon [49] does not currently provide this service in their SLA. The organization would benefit by establishing a cloud-based backup standard, which would clearly define where backup data can be stored, ensuring that the backup is encrypted and performed on a regular basis, as well as ensuring legal and regulatory compliance.

After considering regulatory obligations regarding data retention, current documentation requires that a secure and effective decommissioning process must be defined and implemented into all projects. The methods described in the documents include data-wiping storage media and physically destroying hardware components. This is a potential issue for a cloud-based service for multiple reasons. First, the methods may not be effective in a cloud-based environment or desired by the provider. Second, organizations are unlikely to have access to the physical hardware. Hence, the organization will need to establish with the cloud provider appropriate methods and assurances for the decommissioning of data, retention times and protocols when data breaches occur. These assurances will need to include all infrastructure components along with previous file versions, temporary files and even file fragments.

Ensuring continuous legal and regulatory compliance is a fundamental component of an organization's information security policy [50]. One such example is ensuring compliance with the Data Protection Directive when processing or storing personal data is undertaken [51]. The Directive clearly states that all personal data must be stored inside the European Union (EU) or in 'countries with an adequate level of protection' [51]. To comply with this requirement the organization currently requires that all personal data must be processed within the EU, unless the organization says otherwise. Cloud computing security statements need to be written to include log files, backups conducted by the cloud provider and any other metadata that could be used in incident management for regulatory purposes. Furthermore, the organization should include a security requirement for cloud providers to be able to provide both the logical and physical location of any of the organization's data stored in the provider's infrastructure.

### 4.3 Incident Response

Several security policies and standards targeted the detection and management of security incidents along with subsequent forensic investigations. These documents were created to address investigations involving traditional devices such as servers and personal computers. However, they did not take into consideration cloud computing environments. For example, a policy statement requires that a security analyst begins a chain of custody[2] for any electronic device seized that is suspected to be relevant to an investigation. This is generally an accepted practice in the forensic community [52]. The policy goes on to define how the suspected electronic device can be isolated and disconnected from its power source, allowing the security analyst to seize the device and begin the chain of custody. Creating a chain of custody for cloud computing investigations is anticipated to be more challenging [14, 36]. With data and applications being held by third parties, it is feasible that a chain of custody could be instigated by the cloud provider on behalf of an organization [14]. Due to the plausibility of this scenario, documentation would need to be developed to account for and address appropriate procedures for the creation and management of a chain of custody for cloud environments.

Another potential issue is that the documented forensic imaging methods may be insufficient in cloud environments [14, 36]. Traditionally, a byte-for-byte copy of the entire storage device is made, which is then used for analysis at a later stage [14]. The collection of evidence from cloud environments is unlikely to be straightforward and the tools currently used by the organization to collect potential evidence need to be investigated to determine their effectiveness in a cloud environment. Therefore, the organization needs to develop processes and guidelines to define evidence collection in cloud computing investigations. Documentation will also need to specify what evidence will be collected from cloud services [53]. This will require the creation of provisions to ensure that provider and application logs, along with any other relevant data, are handled appropriately. It is also plausible for organizations to mandate an accurate synchronization of timestamps to aid in the investigation of a security incident.

---
[2] A chain of custody provides a documented history, which should describe how evidence was collected, analysed and preserved in order to be presented as evidence in court (Vacca, 2005).

### 4.4 Encryption and Key Management

Current policies specify that, depending on data classification and regulatory requirements, encryption must be used to achieve data integrity and confidentiality. Policies and guidelines within the ISR define which encryption algorithms and key management processes must be implemented inside the organization. However, what happens when these services are needed outside of the organization? The documents referencing key management need to be expanded to specifically address who is responsible for encryption in a cloud environment. Is this the responsibility of the cloud provider? Regardless of the decision, standards will need to define how encryption keys will be managed between the two entities. Will the cryptographic keys be stored in the cloud itself? Who will have control of the cryptographic keys? One scenario is that policies will require that the organization retain all cryptographic keys within their possession. The drawback to this solution is that it limits the investigative support that is available from the cloud provider. Another issue to take into consideration is the possibility that cloud providers could store its data in countries where authorities have the legal right to request the cryptographic keys from all parties.

### 4.5 Virtualization

The only current documentation which specifically addresses virtualization is a lock-down standard for VMware virtual machines. The organization would need to address three main issues related to virtualization in cloud environments: 'locking down' cloud-based virtual machine instances, mitigating risks from hypervisor-level and cross-virtualization attacks, and virtual machine image security management. The current lock-down document for VMware virtual machines assumes that the administrator enforcing the 'lockdown' has physical access to the virtual machine at the hypervisor level. The standard statements in this document are unlikely to be plausible for a cloud virtual machine instance where the hypervisor is likely to be controlled by the cloud provider. Therefore, an alternative standard needs to be developed to ensure that a virtual machine instance in the cloud is 'locked down' and 'hardened' using policies more suitable for a cloud-based hypervisor. This standard would need to take into consideration the implementation of firewalls for inbound and outbound connections, as well as restrictions on Internet Protocol (IP) addresses for certain levels of access. In addition, this standard would also need to require that the physical machine hosting the hypervisor and virtual machine instance is also 'locked down'. Security controls for applications and data connected to these virtual machine instances also need to be considered and defined.

Security guidelines are also needed to address and mitigate risks associated with hypervisor-level attacks including cross-virtualization attacks [32]. An insecure hypervisor can allow a malicious user to gain access to data stored in virtual machines hosted on a vulnerable hypervisor [32]. Therefore, subsequent requirements need to consider the use of cryptography and isolation as a solution to securing a hypervisor as well as implementing the principle of least privilege [44, 54]. Standards and guidelines will also need to specify how the organization manages a Virtual Machine Image (VMI) [54]. Wei et al., [33] suggest that a framework is developed to manage VMI creation, storage and destruction procedures. This framework is likely to contain controls such as filters to remove sensitive information from an image prior to publishing and a mechanism to track changes performed on a specific image to mitigate malicious image modification [33]. The organization should also ensure that requirements define the secure destruction of decommissioned VMIs.

### 5. Further Work and Conclusions

The increasing popularity of cloud computing services coupled with a growing threat from cyber-criminals creates an environment where security concerns have become a paramount issue. The results of this exploratory case study highlight not only the security issues that are being researched, but also identified opportunities to expand and develop information security documentation. This exploratory research supports the idea that current organizational security policies are potentially putting organizations at risk. It hopefully encourages organizations interested in migrating to the cloud to perform a thorough examination of their security policies, standards and guidelines in reference to cloud environments.

This research presents a foundation for continued examination of cloud security documentation in corporate environments while generating technical and procedural ideas for mitigating future exposure. Future work in this area needs to examine the cloud computing applicant development process, the policies identified for further research and the technical aspects associated with conducting forensic investigations in cloud environments. Research needs to be expanded to examine the most effective way to integrate security into a cloud computing application development process. Research within the organization should conduct a more in-depth examination of the documents and statements that were identified as 'further research required'. Do the technical aspects of these documents and statements present potential security issues for the organization in a cloud environment? Do the organization's current service-level agreements sufficiently protect intellectual property that has been out-sourced to cloud-based service providers? The technical aspects of the tools used to conduct forensic investigations in the cloud need to be investigated. Do the tools that are currently available capture enough data? Ultimately, the case study needs to be expanded to multiple organizations in a variety of industries. The results from an expanded study would provide insight into common problems and encourage research focusing on real-world global solutions. Furthermore, the case study can be repeated in an organization which has already implemented cloud computing. The objective of this new case study would be to determine if the documentation issues identified in the studied organization are actually addressed when a cloud computing environment is implemented.

## 6. Acknowledgments

This work was supported by the A.G. Leventis Foundation. Any opinions, findings, conclusions or recommendations expressed in this paper are those of the authors and do not reflect the views of the A.G. Leventis Foundation. The authors would also like thank the Fortune 500 Organization for their support and feedback in this research